# Unveiling the Scaling Potential of Drain Merge through Active (DMtA) in CFETs: Breaking the Super-Via Bottlenecks and Unlocking New PPA Boosters

Jingru Jiang†, Haoran Lu†, Kairong Guo, Yibo Zhang, Yifei Chen, Wanyue Peng, Yu Liu, Jiacheng Sun, Xiaoyan Xu, Ming Li, Yibo Lin, Runsheng Wang, Ru Huang and Heng Wu*.
†These authors contributed equally, School of Integrated Circuits, Peking University, Beijing 100871, China, *Email: hengwu@pku.edu.cn

***Abstract*****—**Drain merge (DM), a super via vertically connecting the common S/D terminals of stacked n/pFETs in Complementary FETs (CFETs), blocks further parasitic optimization and cell scaling. For the first time, this work systematically investigates the state-of-the-art Drain Merge through Active (DMtA), a revolutionary technology reported recently with the DM embedded in the active region, through a comprehensive DTCO framework spanning process integration, contact-configuration-dependent (CTCD) compact modeling, standard-cell design, RO evaluation and block-level PPA benchmark on a 32-bit RISC-V Ibex core. By reducing DM parasitics and enabling DM-width optimization, DMtA improves RO frequency by 11.7% over its conventional Drain Merge through field (DMtF) counterpart. ***Active widening and Area Borrowing***, the latter first reported in [8] and exploiting spatial slack in adjacent cells to further enlarge the nanosheet width ($W_{NS}$), increase the maximum Ibex-core frequency by up to 34.8%. More importantly, DMtA also enables the ***once GAA-exclusive Hyper-cells*** on CFETs by merging the active regions across adjacent cell rows, providing a further 8.7% frequency gain. A post-routing floating-output-pin-aware optimization further ***removes redundant S/D contacts*** (CTs) and reduces power by 5.3%. Finally, DMtA facilitates ***more area-efficient 2.5T cell scaling*** by preserving single-row cell compatibility, reducing post-PR core area by 25.7%.

## INTRODUCTION

Continued logic scaling increasingly relies on vertically stacked transistors, with CFETs [1-3] offering reduced cell footprints and enhanced performance. However, their benefits are inevitably limited by super vias connecting the stacked device tiers [4,5]. In conventional CFETs, DMtF [5-7], a super via located outside the active region that connects the common S/D terminals of stacked n/pFETs, introduces parasitic and geometric limits [4,5] that hinder performance improvement and standard-cell scaling. DMtA [1-3] has been proposed to mitigate these limitations by embedding the DM in the active region, where it passes through the top S/D epi and lands directly on the bottom S/D epi (Figs. 1(a-b)). Reported as epi-to-epi via (EEV) by Intel [1] (Fig. 1(c)), Thru-contact by Samsung [2] and vMDLI [3] by TSMC (Fig. 1(d)), DMtA has shown potential to alleviate DM's RC penalty [1,2]. However, its potential for PPA improvement in CFETs remains largely unexplored.

In this work, DMtA, as a paradigm-shifting DTCO knob for CFETs, is comprehensively investigated from device to chip levels, with a broader set of PPA boosters proposed, as in Fig. 2. By releasing the field-region space, DMtA enlarges $W_{NS}$ and DM width ($W_{DM}$), which can be further extended through Area Borrowing [8] using spatial slack from adjacent cells. It also enables active merging across adjacent cells to form Hyper-cells [9], jointly improving performance. Meanwhile, redundant CT removal reduces capacitance and power, while the relaxed DM-spacing constraint enables further cell-height scaling from 3T to 2.5T [5-7]. In this work, these boosters are systematically examined from process integration and compact modeling to standard-cell, RO, and block-level evaluations. Scalability is further demonstrated through 2.5T cell scaling and extension to CFET-based FFET (CFFET) [10] with 4-tier transistors enabled by back-to-back-stacked CFETs.

## COMPACT MODEL AND EVALUATION FRAMEWORK

### *A. DMtA Process and Compact Modeling*

The CFET in this work adopts an FS-BS symmetric architecture and follows the double-flip FFET [4,10] integration flow reported in [10], as shown in Fig. 3. After self-aligned active and dummy-gate patterning, the frontside (FS) n-S/D epi is formed, followed by the first wafer flipping and backside (BS) pFET fabrication. After the second wafer flipping, a dedicated DMtA trench is etched through the FS n-S/D epi and lands on the BS p-S/D epi. Subsequent metal fill forms the DMtA together with the FS CTs, followed by other FS BEOL processes.

Besides the unique process requirement, DMtA also introduces a compact-modeling challenge by modifying the local S/D epi and CT morphology, resulting in CTCD device characteristics. Taking the DFQD1 cell as an example, Fig. 4(a) identifies 4 CT types: a conventional CT on top of the S/D epi (Type-1), a vertical CT through the DMtA-penetrated S/D epi (Type-2), a bottom CT formed at the S/D epi on which the DMtA lands (Type-3), and a combined top-and-bottom CT when Types-1 and Type-3 coexist (Type-4). Their corresponding S&D terminals are color-coded in the DFQD1 schematic. Assigning the 4 CT types to S&D terminals yields 10 symmetry-distinct S&D terminal configurations, whose TCAD $I_dV_g$ characteristics are shown in Fig. 4 (b), with the corresponding device assumptions summarized in Table 1. The pFET variants exhibit a normalized $I_{dsat}$ range of 10.5% at $V_{dd}$ = 0.7 V, indicating an obvious CTCD variation.

To evaluate the circuit-level impact, Fig. 4(c) considers a transmission gate (TG) with Type-2 CTs for both pFET S&D terminals and Type-3 CTs for both nFET S&D terminals and defines the propagation delay ($t_{pLH}$) from the bottom output waveform. Compared with the conventional approach using models with Type-1 CTs for S&D terminals, the proposed CTCD modeling results in a 6.9% difference in $t_{pLH}$ (Fig. 4(d)), highlighting the importance of capturing DMtA-specific CT configurations for accurate circuit evaluation.

### *B. Evaluation Framework*

A 15-stage ring oscillator (RO) with fan-out of 3 (no BEOL loading) was used for circuit-level comparisons. Unless otherwise specified, all RO evaluations in this work adopted this topology. Block-level evaluations were performed using a 32-bit RISC-V Ibex core [11] (only computing core, no cache), implemented with DMtF and DMtA CFET libraries under A7 technology assumptions. The corresponding design rules are listed in Tables 1 and 2. Note that the CFET std. cells adopt layouts with dual-sided (DS) pins [4,7,12,13], enabling DS signal routing [12,14] during physical implementation. The evaluated CFET standard cell libraries comprise over 2000 cells across all considered technology and design options, and were characterized at $V_{dd}$ = 0.7 V.

## DMTA-ENABLED PPA BOOSTERS

### *A. Initial Evaluation and DM-Width Optimization*

Unlike the field-region DMtF, whose width is constrained by the required spacing to the active and the adjacent DMtF, DMtA provides greater flexibility for $W_{DM}$ optimization (Fig. 5(a)) and DM parasitic reduction. With both structures using baseline $W_{DM}$ = 10 nm and the W1 active-width option of $W_{NS}$ = 14 nm, 3T DMtA CFET achieves an 8.2% INVD1-based RO frequency gain at $V_{dd}$ = 0.7 V over the 3T DMtF CFET (Fig. 5(b)), owing to 33.8% lower n-drain-to-p-drain resistance ($R_{nd\text{-}pd}$) and 16.9% lower input-to-output capacitance ($C_{in\text{-}out}$) (Fig. 5(c)); the resistance reduction stems from the reduced DM height ($H_{DM}$), while the capacitance reduction results from eliminating the gate-overlapping portions of the field-region CT and DM (Fig. 5(a)). Leveraging the greater $W_{DM}$ tunability of DMtA, $W_{DM}$ was further swept from 10 to 22 nm. At fixed $W_{DM}$ = 16 nm, the $R_{nd\text{-}pd}$ reduction over DMtF CFET increases to 60.5%, yielding the peak RO frequency with an 11.7% gain over DMtF CFET (Figs. 5(b-c)).

Based on the optimized $W_{DM}$ of DMtA, 3T DMtF and DMtA CFET libraries were evaluated on the Ibex core. The physical design results, including DS signal routing, is shown in the post-PR layouts in Fig. 6(a). The power–frequency curves in Fig. 6(b) show that the DMtA CFET achieves up to 7.1% higher frequency at iso-power and up to 16.4% lower power at iso-frequency than the DMtF CFET. Thus, the intrinsic DM parasitic advantage of DMtA already translates into measurable block-level performance and power benefits before further DTCO.

### *B. Active Widening and Area Borrowing*

DMtA further enables active widening by releasing the field-region space originally occupied by the field-region DMtF. As shown in Fig. 7(a), this structural advantage directly expands the $W_{NS}$ from W1 = 14 nm in the 3T DMtF CFET to W2 = 26 nm in the 3T DMtA CFET. Building on W2 cells, Area Borrowing [8] further enlarges the $W_{NS}$ to W3 = 36 nm by utilizing the spatial slack of an adjacent W1 cell, while maintaining valid W3-W1 cell abutment, as shown in Fig. 7(b). The progressive increase in effective current ($I_{eff}$) with $W_{NS}$ translates into 35.7% and 21.5% higher INVD1-based RO frequencies for DMtA W2 over W1 and W3 over W2, respectively, at $V_{dd}$ = 0.7 V (Fig. 7(c)).

At the block level, W2 cells provide a higher effective channel width ($W_{eff}$) at the same footprint, enabling the area-saving switching of high-drive (D2/D4/D8) W1 cells to smaller-area D1 W2 cells. This reduces the high-drive cell count by 73.2% and the total cell count by 15.6%, resulting in a 16.9% reduction in total cell area (Figs. 8(a-b)).

During the placement and clock-tree synthesis, a limited number of W3 cells were used to effectively improve timing. Fig. 9(a) shows W3 cell usage increases with achieved frequency, indicating their growing role in timing optimization, yet remains below 2% of the total cell count, leaving sufficient space for cell abutment constraint and ensuring Area Borrowing during legalization. Static timing analysis further shows that the top-100 to top-1000 critical paths (CPs) contain more than four W3 cells per path on average and every analyzed critical path includes at least one W3 cell (W3 path coverage = 100%) (Fig. 9(b)). This proves Area Borrowing a highly efficient performance booster.

Core utilization was swept to map frequency vs core area and determine the max. frequency ($f_{max}$). Introducing W2 cells increases $f_{max}$ by 27.8% over the W1-only library, while further enabling W3 cells extends the gain to 34.8%, confirming the progressive benefits of active widening and Area Borrowing.

### *C. Active Merging with Hyper-cells*

Hyper-cells have recently been proposed in GAA technologies to enhance cell performance by merging the active regions of parallel transistors across rows [9]. However, this technology cannot be directly applied to DMtF CFETs because the DMtF obstructs active merging (Fig. 11(a)). With the DM embedded in the active, Hyper-cells become feasible in DMtA CFETs because the obstruction to active merging is eliminated (Fig. 11(b)). The corresponding layouts illustrate how merging the active regions of two adjacent 3T W2 cells forms a 6T double-row Hyper-cell, increasing $W_{eff}$ by 66.1% (Fig. 11(c)). For an iso-area RO comparison, the Hyper INVD1 (2CPP × 2Row) and W2 INVD3 (4CPP × 1Row) were evaluated, with the former achieving a 43.5% higher frequency at iso-power and 58.9% lower power at iso-frequency relative to the latter at $V_{dd}$ = 0.7 V, resulting in a 53.1% reduction in energy-delay product (EDP) at iso-leakage. For the block-level assessment, a baseline library containing W1 and W2 cells was compared to an extended library with additional Hyper-cells. The design with Hyper-cells achieves an 8.7% higher $f_{max}$, 4.1% lower power, and 9.6% lower total cell area at their respective $f_{max}$ points (Figs. 12(a-b)).

Fig. 13 further highlights that Hyper-cells can be more efficiently implemented in the DMtA CFET than in GAA. As shown in Fig. 13(a), GAA-based Hyper-cells support only N-active merging or P-active merging, whereas the vertically stacked active in the DMtA CFET enables simultaneous N/P active merging, yielding $W_{eff}$ gains of 54% and 108% (Fig. 13(b)), respectively, and frequency gains of 17%, 28% and 56% (Fig. 13(c)) relative to their unmerged counterparts.

### *D. Redundant S/D Contact Removal and Design Optimization*

By landing directly on the opposite-side S/D epi rather than the opposite-side CT, DMtA—unlike DMtF—enables removal of the redundant CT when output access from the opposite side is not required. Figs. 14(a-b) show that FS-/BS-formed DMtA (FS/BS DMtA) enables removal of the redundant BS/FS CT, respectively, reducing the drain capacitance ($C_{drain}$) by 13%. To evaluate this benefit, a dedicated 101-stage RO was formed by alternately cascading BS-in/FS-out (BIFO) and FS-in/BS-out (FIBO) INVD1 cells (Fig. 15(a)). Fig. 15(b) compares FS DMtA without CT removal, FS DMtA with CT removal only enabled in FIBO, and DS (FS+BS) DMtA with CT removal in both BIFO and FIBO; the latter two increase RO frequency by 4.8% and a further 2.7%, respectively, as DS DMtA removes redundant CTs from both sides and thus achieves the lowest effective capacitance ($C_{eff}$) (Fig. 15(c)).

More generally, although DS DMtA enables redundant CT removal for either FS or BS output access, backend optimization is required to fully realize its chip-level benefits. Fig. 16(a) illustrates the basic DS signal routing [12,14] enabled by DM-connected DS output (DO) pins in CFET cells [6,7,13], allowing the FS or BS output pin to be selected for cascading according to the input-pin side of the next stage [6,12,14], therefore forming FS-only, BS-only, or DS nets. When a DO pin drives a FS-/BS-only net, the unused-side pin floats and can be removed together with its redundant CT by converting the original DO cell into an FS-only-output (FO) or BS-only-output (BO) cell (Fig. 16(b)). A floating-output-pin-aware post-PR cell optimization workflow was developed to automatically realize this DO-to-FO/BO conversion as shown in Fig. 17. Applying this workflow to a 3T-DMtA-CFET-based Ibex core converts 55.4% of DO cells to FO/BO cells and reduces total power by 2.5% (Fig. 18(a)), primarily through lower internal power enabled by redundant CT removal (Fig. 18(b)).

The remaining DO cells in Fig. 18(a) show that further DO-to-FO/BO conversion requires reducing DS-net usage, which otherwise limits redundant-CT removal. This can be addressed by updating cell input configurations [13-15] by relocating input pins between the FS and BS, consequently converting DS nets into FS- or BS-only nets [14,15] and enabling additional DO-to-FO/BO conversions (Fig. 19(a)). Broader input-configuration coverage [13]—the ratio of total available to theoretically possible input configurations—reduces DS nets by up to 65.8%, enabling more DO-to-FO/BO conversions and increasing the power saving from redundant CT removal by up to 29.1% at 2.5 GHz (Fig. 19(b)). This ultimately yields a 5.3% total power reduction relative to the design before redundant CT removal.

Fig. 20 summarizes the cumulative Ibex-core frequency gains enabled by 3T DMtA CFET optimizations over the 3T DMtF baseline. Building on NS widening and Hyper-cells, redundant CT removal raises the frequency gain to 26.2% at iso-power = 5.4 mW, while broader input-configuration coverage further increases it to 29.3%.

### *E. Area-Efficient Scaling to 2.5T with DMtA*

At the 2.5T cell height, DMtFs in adjacent rows violate the min. DM-spacing rule, forcing less area-efficient double-row cells [16], whereas DMtA avoids this conflict and preserves single-row cells (Fig. 21(a)). Even within a double-row cell, DMtA eliminates the lateral staggering required by DMtF and thus reduces cell width—for example, by 2 CPP for DFQD1 (Fig. 21(b)). These advantages translate into an average library area reduction of 12.0% (Fig. 21(c)) and, together with a 9.5% lower cell number, a 25.7% post-PR Ibex-core area reduction (Fig. 21(d)) for the 2.5T DMtA CFET relative to the 2.5T DMtF double-row CFET.

### *F. Extension to 4-Tier CFFETs*

DMtA can also serve as a performance booster for 4-tier-transistor CFFETs [10]. In the CFFET, the aligned-active arrangement requires VHV routing [10,17] to bring internal signals to the M0 tracks above the active region, increasing the MOL vertical-interconnect height ($\Delta H_{MOL}$) and resistance (Fig. 22(a)). Staggering the active regions [18] enables direct vertical connection without VHV. For staggered-active CFFETs, DMtA avoids the cell-boundary extension required by DMtF and the associated area penalty (Fig. 22(b)). Relative to the aligned-active DMtF CFFET, the aligned-active and staggered-active DMtA CFFETs improve RO frequency by 9.7% and 22.8%, respectively, with the latter benefiting from reduced MOL resistance, as shown in Fig. 22(c).

## CONCLUSIONS

In this work, for the first time, we fully investigated the novel DMtA architecture on CFET from process integration and compact modeling to circuit- and block-level implementations. By intrinsically reducing DM resistance and parasitic capacitance while further enabling active widening, Area Borrowing, Hyper-cells, redundant CT removal, and area-efficient 2.5T cell scaling newly proposed on CFETs, DMtA is validated through comprehensive PPA evaluation as a scalable DTCO solution for future CFET technologies.

## ACKNOWLEDGMENT

This work was supported in part by the National Key R&D Program of China under Grant 2023YFB4402201; in part by the NSFC under Grant 92464206; and in part by GJ Program under Grant 202504841098.

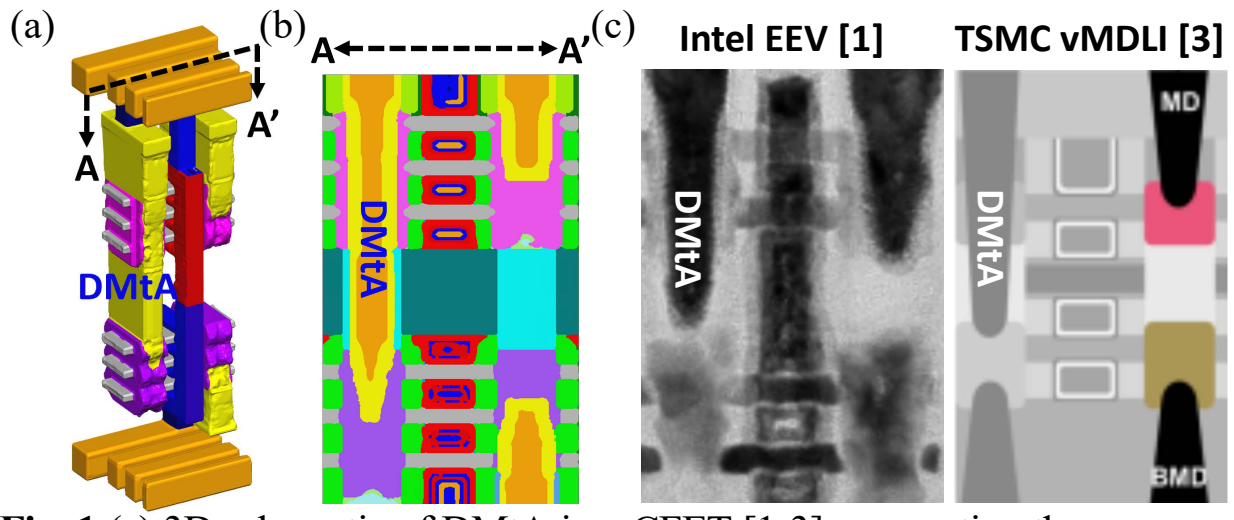

**Fig. 1** (a) 3D schematic of DMtA in a CFET [1-3], connecting the common S/D terminals of stacked n/pFETs through the top S/D epi to the bottom S/D epi. (b) Cross-gate cartoon of the DMtA. (c) TEM image of Intel's DMtA process, termed EEV [1]. (d) TSMC's DMtA structure, termed vMDLI [3].

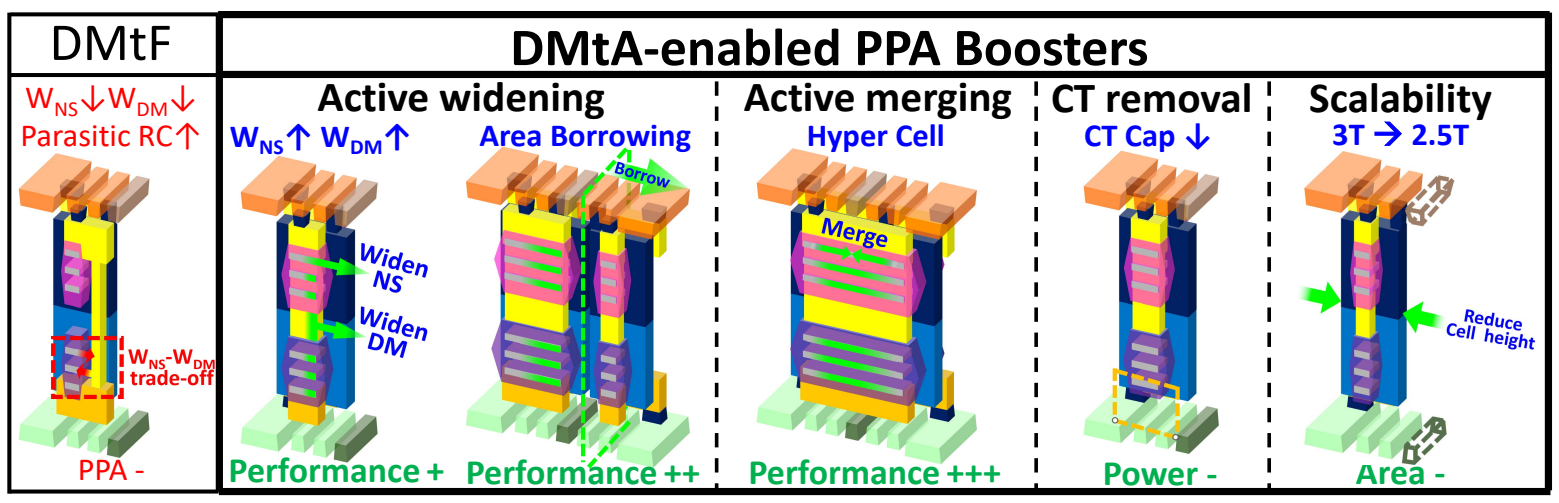

**Fig. 2** DMtF reference and the unique DMtA-enabled PPA boosters. DMtA expands active and DM width, while Area Borrowing exploits spatial slack from adjacent cells; active merging adjacent cells to form Hyper-cells, jointly enhancing performance. Redundant CT removal lowers capacitance and power, while DMtA enables further cell-height scaling from 3T down to 2.5T.

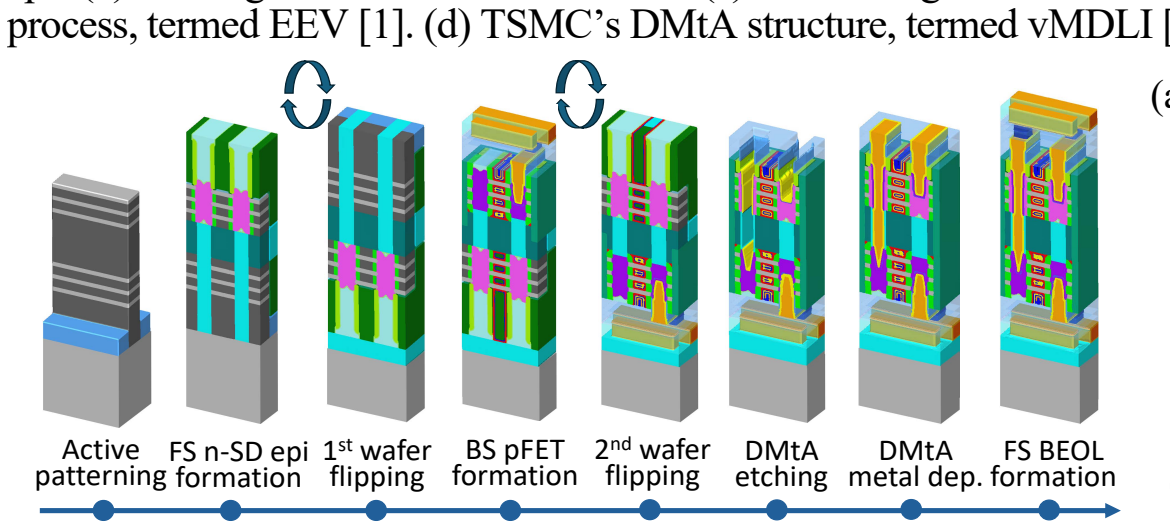

**Fig. 3** The overall integration of the FS-BS symmetric CFET follows the double-flip FFET [4,10] flow reported in [10]. During CT formation, a dedicated DMtA trench is etched within the active region through the FS n-S/D epi and lands on the BS p-S/D epi, followed by contact formation and metal deposition.

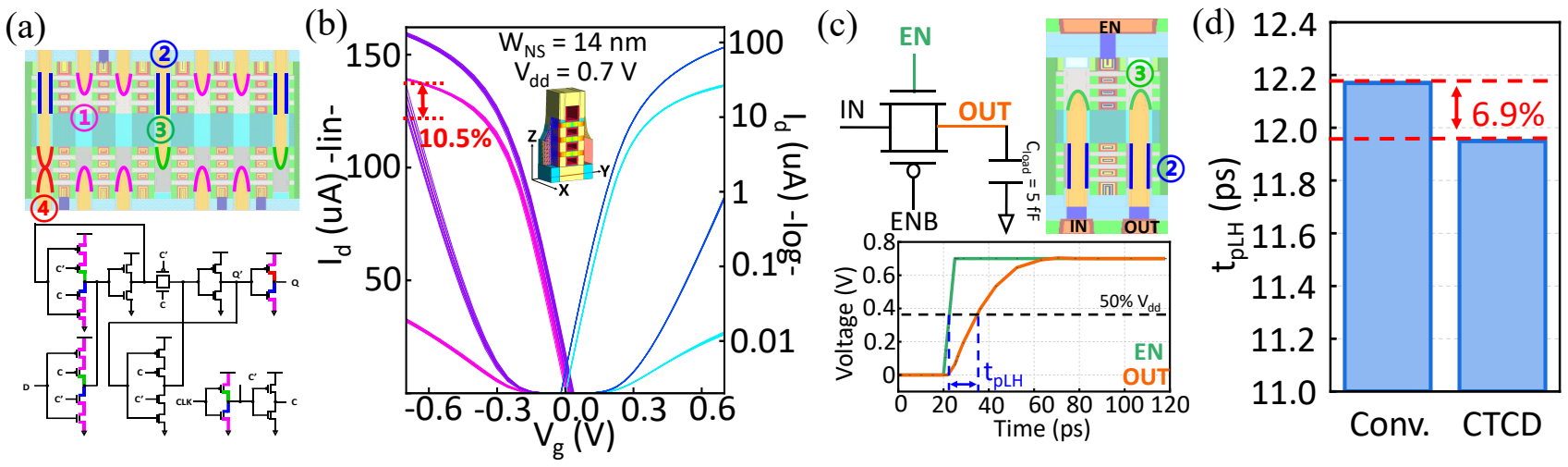

**Fig. 4** CTCD compact modeling for DMtA. (a) 4 CT types and their assignments in a CFET DFQD1 cell. (b) TCAD $I_d$–$V_g$ characteristics of nFET and pFET variants for the 10 symmetry-distinct S&D terminal configurations. (c) TG benchmark comparing the conventional model with Type-1 CTs for S&D with CTCD models (p-S&D: Type-2; n-S&D: Type-3), with $t_{pLH}$ defined. (d) Resulting 6.9% $t_{pLH}$ difference.

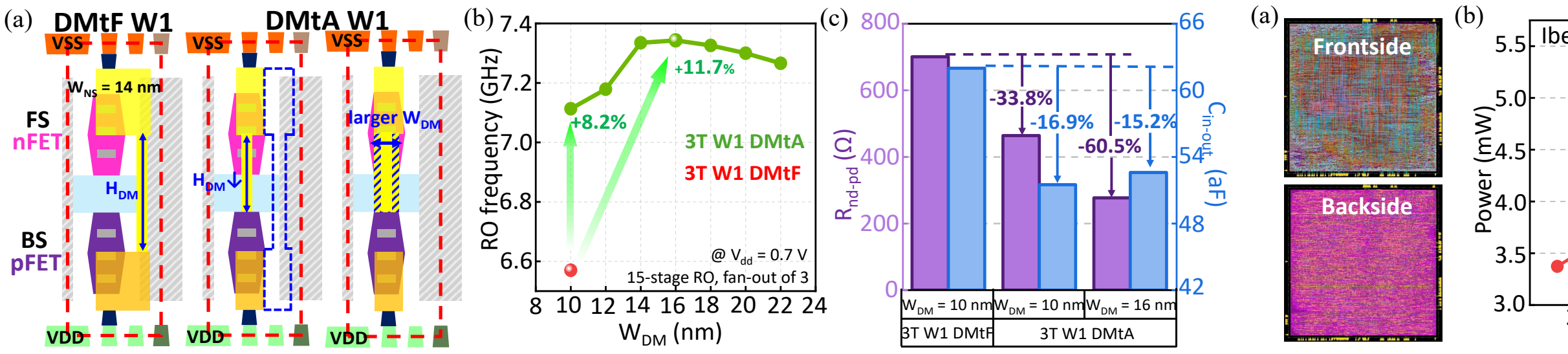

**Fig. 5** (a) Schematics of 3T DMtF CFET (left), 3T DMtA CFET with the same $W_{DM}$ (middle), and 3T DMtA CFET with an enlarged DM (right). (b) INVD1-based RO frequency vs $W_{DM}$. (c) $R_{nd\text{-}pd}$ and $C_{in\text{-}out}$ of 3T DMtF CFET and 3T DMtA CFET with 10- and 16-nm-wide DMs, revealing the RC trade-off as the DMtA extends beyond the active region.

**Fig. 6** (a) Post-PR layouts of a 3T CFET Ibex core with DS signal routing. (b) Power-frequency curves of the 3T DMtF CFET and 3T DMtA CFET (Ibex core).

| CFET Architecture | DMtF CFET | | DMtA CFET | |
|---|---|---|---|---|
| Std. Cell Track # | 3T | 2.5T | 3T | 2.5T |
| Contact Poly Pitch (nm) | 45 | | | |
| Gate Length (nm) | 15 | | | |
| Nanosheet Width (nm) | 14 | 14 | 14/26/36/98 | 14 |
| Nanosheet Thickness (nm) | 5 | | | |
| Sheet-to-sheet Spacing (nm) | 10 | | | |
| EOT (nm) | 0.83 | | | |
| Gate Spacer Thickness (nm) | 5.5 | | | |
| nFET Channel & EPI | Si & Si:P | | | |
| pFET channel & EPI | $Si_{0.65}Ge_{0.35}$ & $Si_{0.3}Ge_{0.7}$:B | | | |
| N-P Vertical Spacing (nm) | 40 | | | |
| DM Width (nm) | 10 | 10 | 16/28/38/100 | 16 |
| Cell Height | 72 | *120 | 72 | 60 |

* Double-row cell

**Table 1** Device assumptions and std. cell design rules for the A7 CFETs in this work.

| Layers | Pitch (nm) | CD (nm) |
|---|---|---|
| FM11-FM10 | 126 | 62 |
| FM9-FM7 | 76 | 38 |
| FM6 | 42 | 20 |
| FM5-FM4 | 35 | 18 |
| FM3 | 32 | 16 |
| FM2 | 24 | 14 |
| FM1 | 45 | 16 |
| FM0 | 20 | 12 |
| BM0 | 20 | 12 |
| BM1 | 45 | 16 |
| BM2 | 24 | 14 |
| BM3 | 32 | 16 |
| BM4-BM5 | 35 | 18 |
| BM6 | 42 | 20 |
| BM7-BM9 | 76 | 38 |
| BM10-BM11 | 126 | 62 |
| BM12-BM13 | 950 | 430 |

**Table 2** Design rules of the BEOL metal layers.

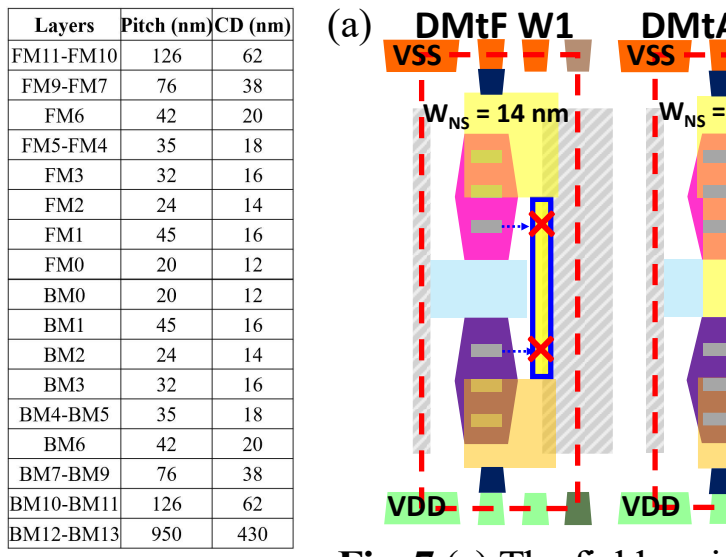

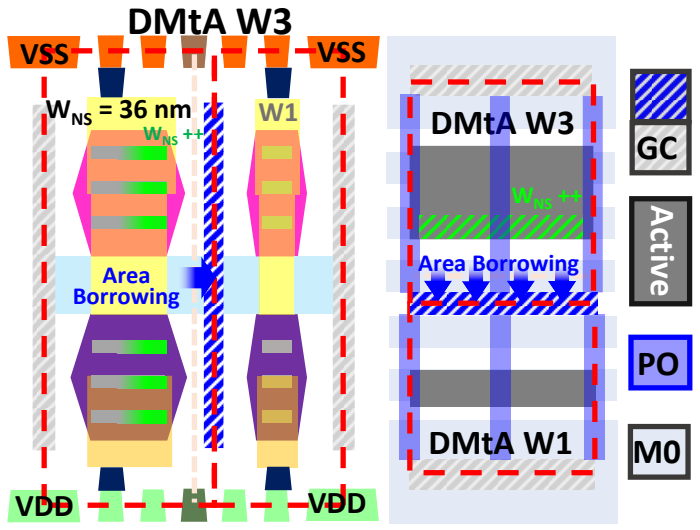

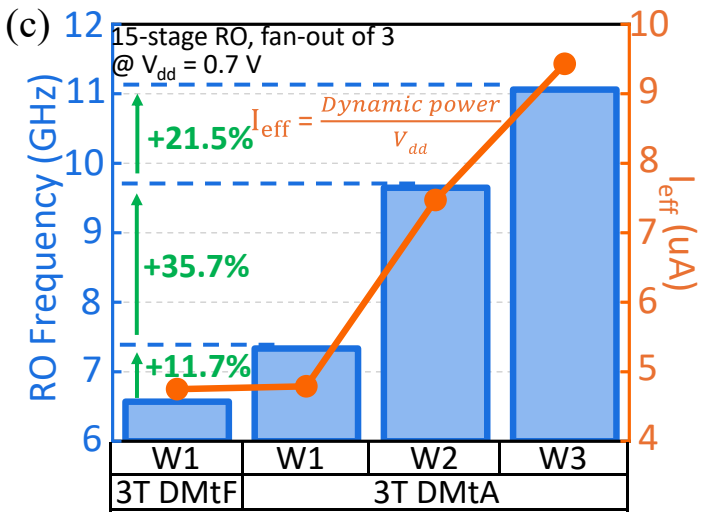

**Fig. 7** (a) The field-region DMtF limits $W_{NS}$ to W1 = 14 nm, while the DMtA releases the field-region space and expands $W_{NS}$ to W2 = 26 nm. (b) Area Borrowing [8], enabled by DMtA, expands $W_{NS}$ to W3 = 36 nm by utilizing the spatial slack from an adjacent W1 cell. The right layout shows valid W3-W1 cell abutment. (c) INVD1-based RO (Fan-out = 3, stage = 15) frequency and $I_{eff}$ at $V_{dd}$ = 0.7 V for 3T DMtF CFET and 3T DMtA W1/W2/W3 CFET cells.

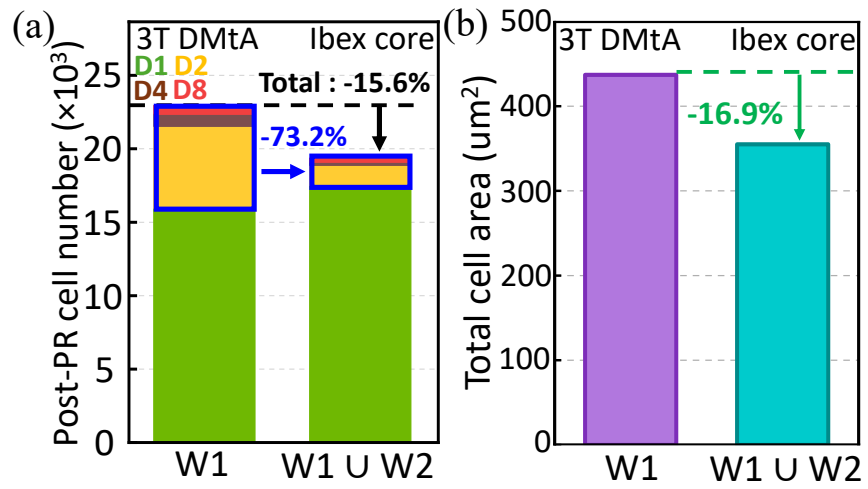

**Fig. 8** (a) Post-PR (Ibex core) cell distribution for W1-only and W1∪W2 3T DMtA libraries. Introducing W2 cells reduces the use of high-drive (D2/D4/D8) cells, thereby lowering total cell area in (b).

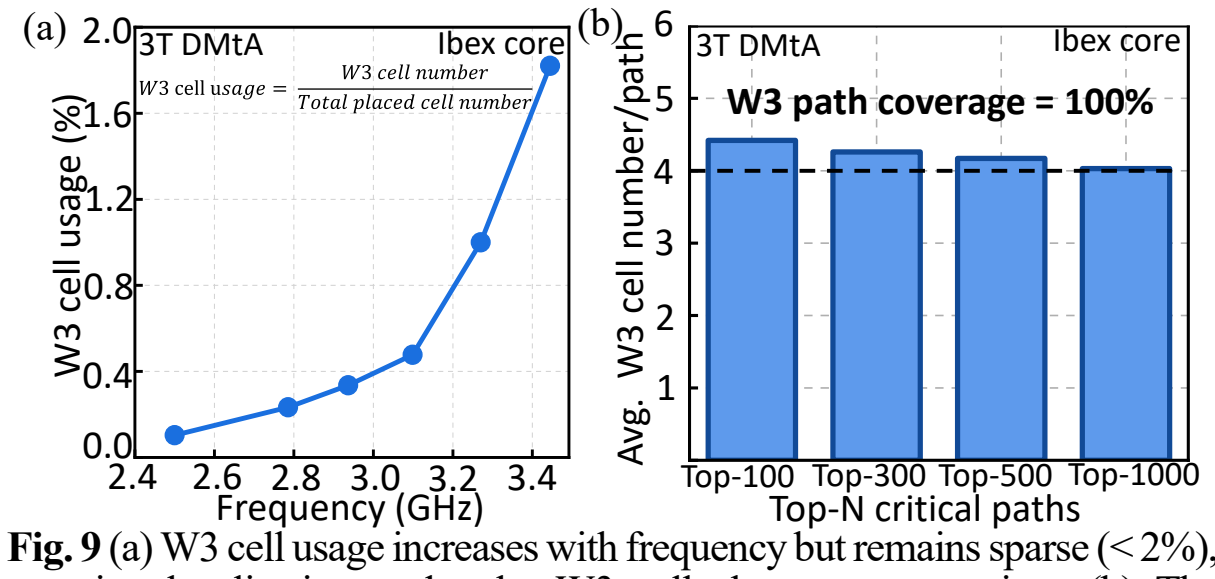

**Fig. 9** (a) W3 cell usage increases with frequency but remains sparse (< 2%), ensuring legalization under the W3 cell abutment constraints. (b) The average W3-cell count exceeds four per path across the top-100 to top-1000 CPs, with 100% path coverage, indicating intensive W3-cell usage in CPs.

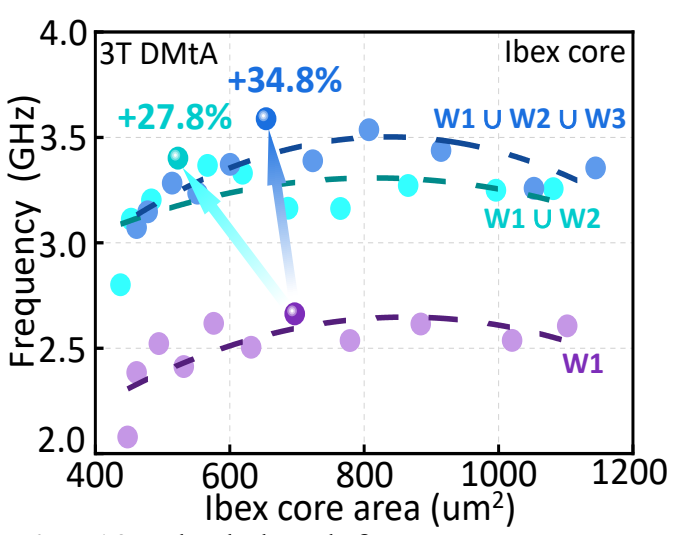

**Fig. 10** Block-level frequency vs core area for W1-only, W1∪W2 and W1∪W2∪W3 3T DMtA libraries, showing a progressive increase in $f_{max}$.

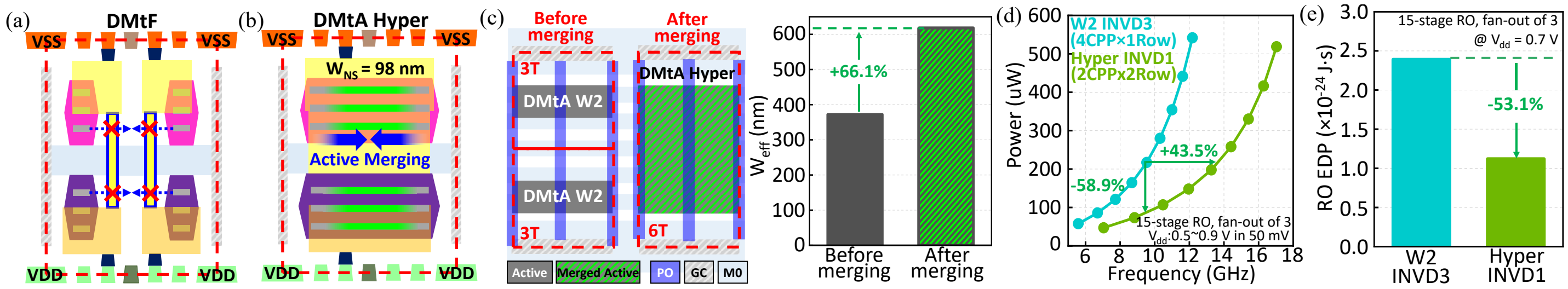

**Fig. 11** (a) DMtF obstructs active merging between adjacent cells. (b) Without the obstruction, DMtA enables the Hyper-cell [9] in CFETs. (c) Taking the 3T W2-cell layout as an example, parallel-transistor active is merged to form the 6T Hyper-cell layout (left), substantially increases $W_{eff}$ over the W2 cell (right). (d) RO (fan-out = 3, stage = 15) power-frequency relation of the W2 INVD3 and Hyper INVD1 with the same cell area, validating improved power and frequency in Hyper-cells. (e) The resulting EDP reduction.

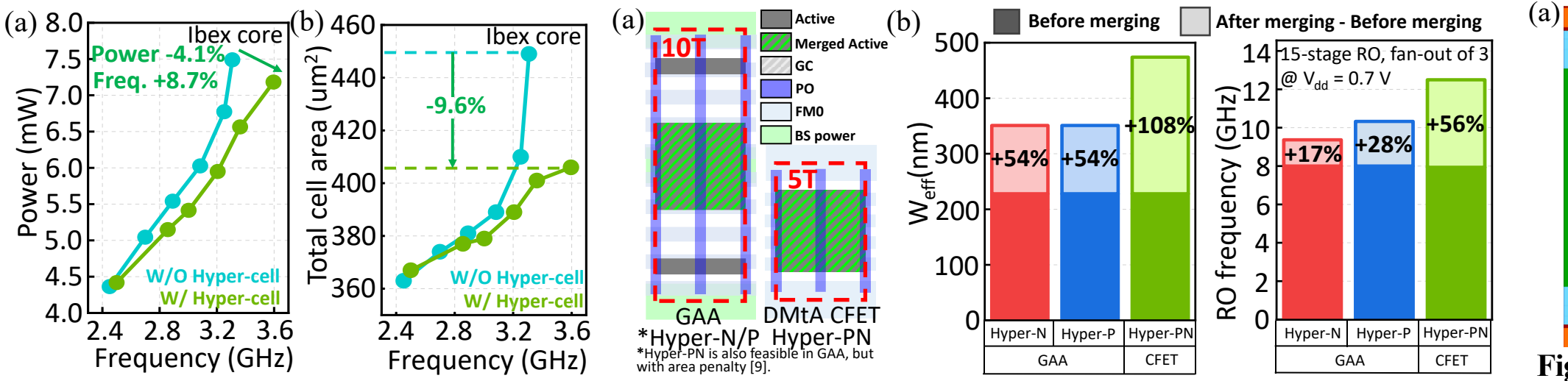

**Fig. 12** Hyper-cells enable 3T DMtA CFET Ibex core (a) 8.7% higher max. frequency and 4.1% lower power, and (b) 9.6% less total cell area, all studied at the respective max frequency.

**Fig. 13** (a) Layouts of 10T GAA-based and 5T CFET-based Hyper-cells. CFET's stacked active enables N/P active merging together, whereas GAA permits only N- or P-active merging. (b) Active merging in CFET yields a larger $W_{eff}$ gain (left) and thus a higher frequency gain (right).

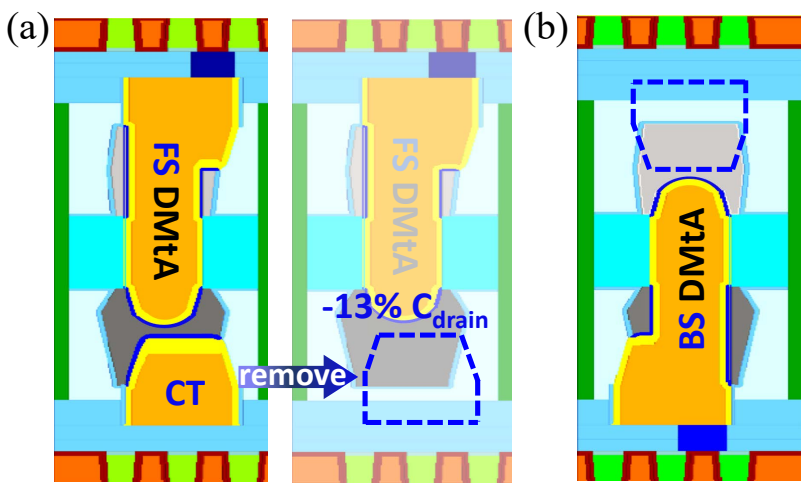

**Fig. 14** (a) Landing on the BS S/D epi instead of CTs, DMtA formed on FS can remove redundant BS contacts, reducing $C_{drain}$ by 13%. (b) DMtA formed on BS can remove redundant FS contacts.

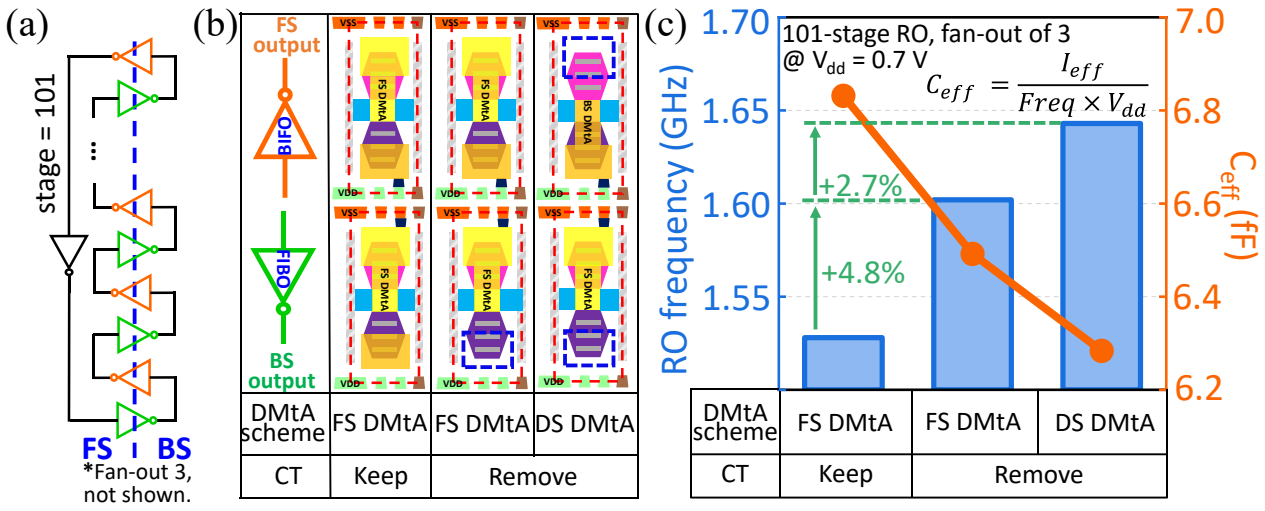

**Fig. 15** Dedicated RO evaluation of FS and DS DMtA for redundant CT removal. (a) RO formed by alternately cascading BS-in/FS-out (BIFO) and FS-in/BS-out (FIBO) INVD1. (b) INVD1 structures for the 3 designs: FS DMtA without CT removal; FS DMtA with FIBO-only removal, and DS (FS+BS) DMtA with CT removal in both BIFO and FIBO. (c) RO frequency and $C_{eff}$.

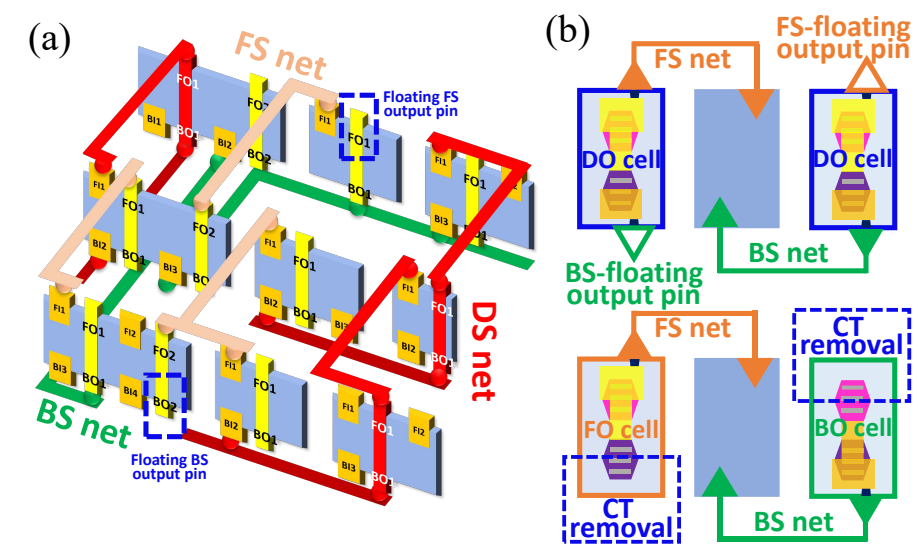

**Fig. 16** (a) Illustration of DS routing on DS output pins by DM. (b) Not all DS output pin drives both FS & BS nets, leading to FS- or BS-floating output pins, can be removed to eliminates the redundant CTs.

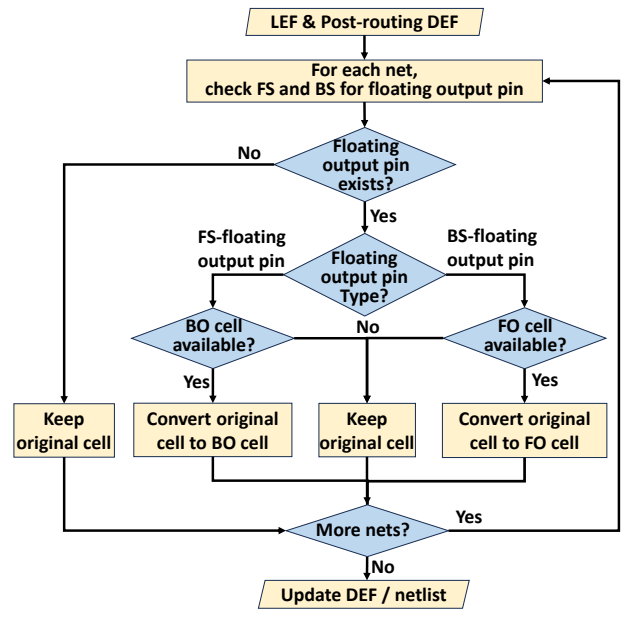

**Fig. 17** Workflow of floating-output-pin-aware post-PR cell optimization.

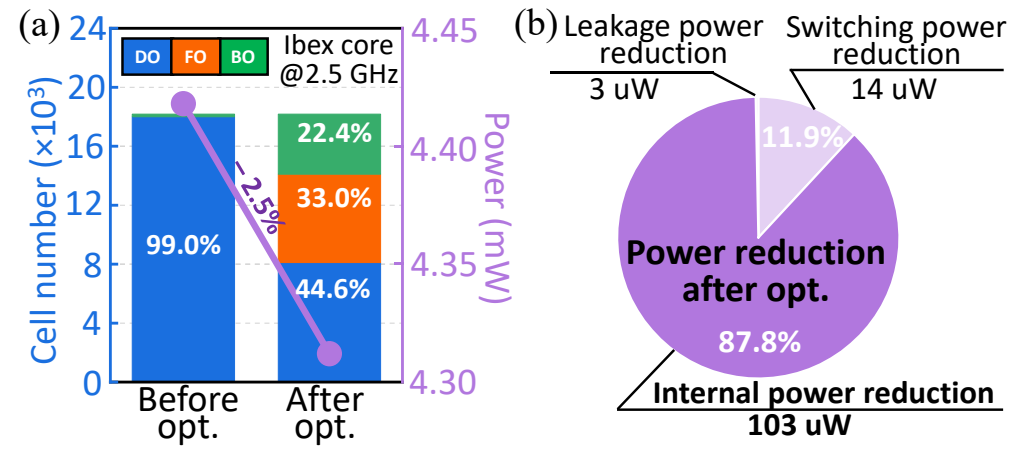

**Fig. 18** DO-to-FO/BO cell optimization using a 3T DMtA CFET library. (a) Converting 55.4% of DO cells to FO/BO cells delivers a 2.5% power saving, mainly owing to (b) lower internal power enabled by the redundant CT removal.

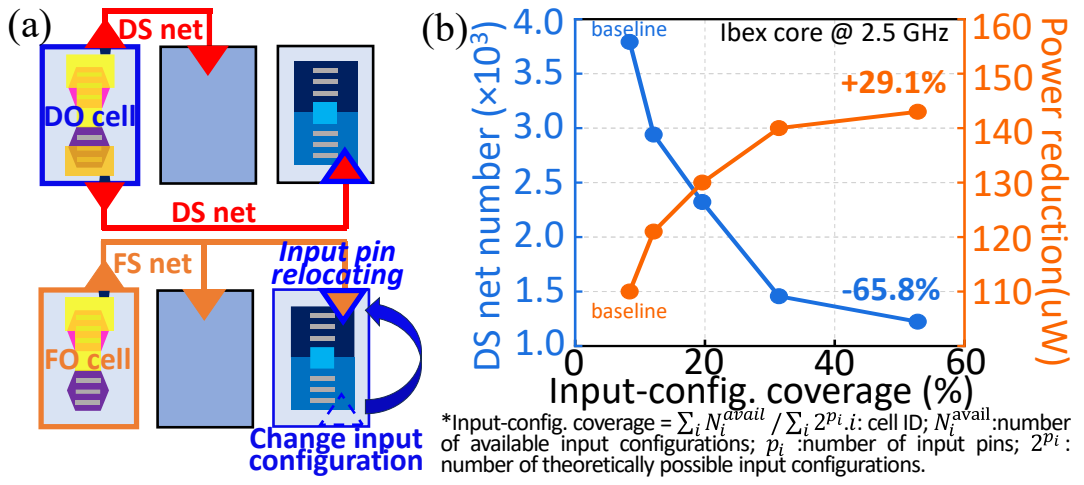

**Fig. 19** Input-configuration optimization for redundant CT removal. (a) FS/BS input-pin relocation converts DS nets into single-sided nets, enabling additional DO-to-FO/BO conversions. (b) Broader input-configuration coverage reduces DS nets and increases power savings.

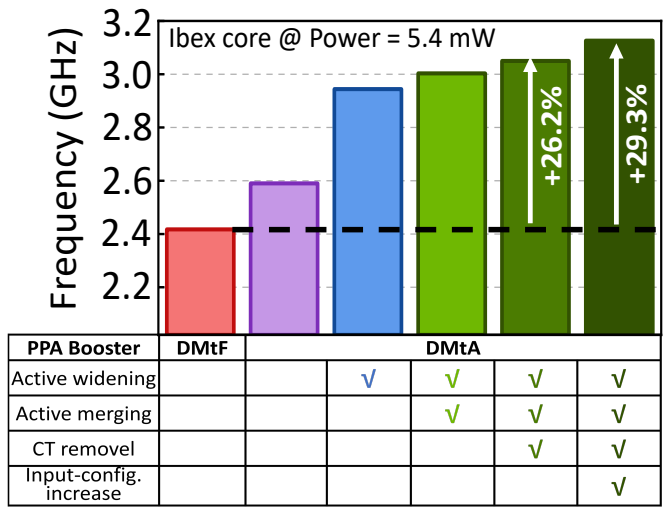

**Fig. 20** Summary of Ibex-core frequency gains enabled by 3T DMtA CFET optimizations over the 3T DMtF CFET.

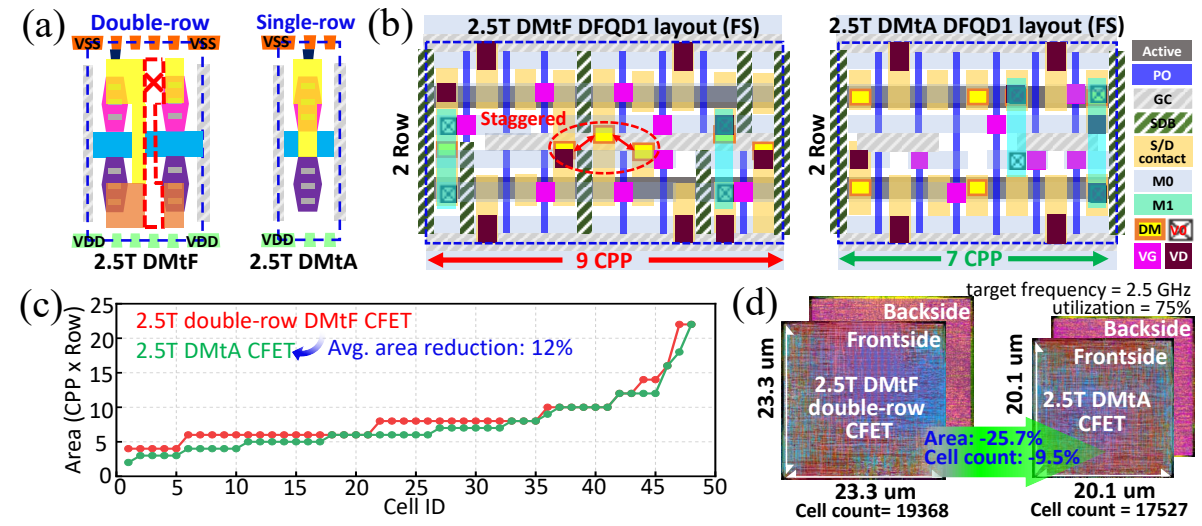

**Fig. 21** Area-efficient scaling to 2.5T with DMtA. (a) DMtFs in adjacent rows violate min. DM spacing and requires less area-efficient double-row cells, whereas DMtA retains single-row cells. (b) DMtF requires lateral staggering, incurring a 2-CPP width penalty for DFQD1. (c,d) Library and core area comparisons between 2.5T double-row DMtF and DMtA CFETs.

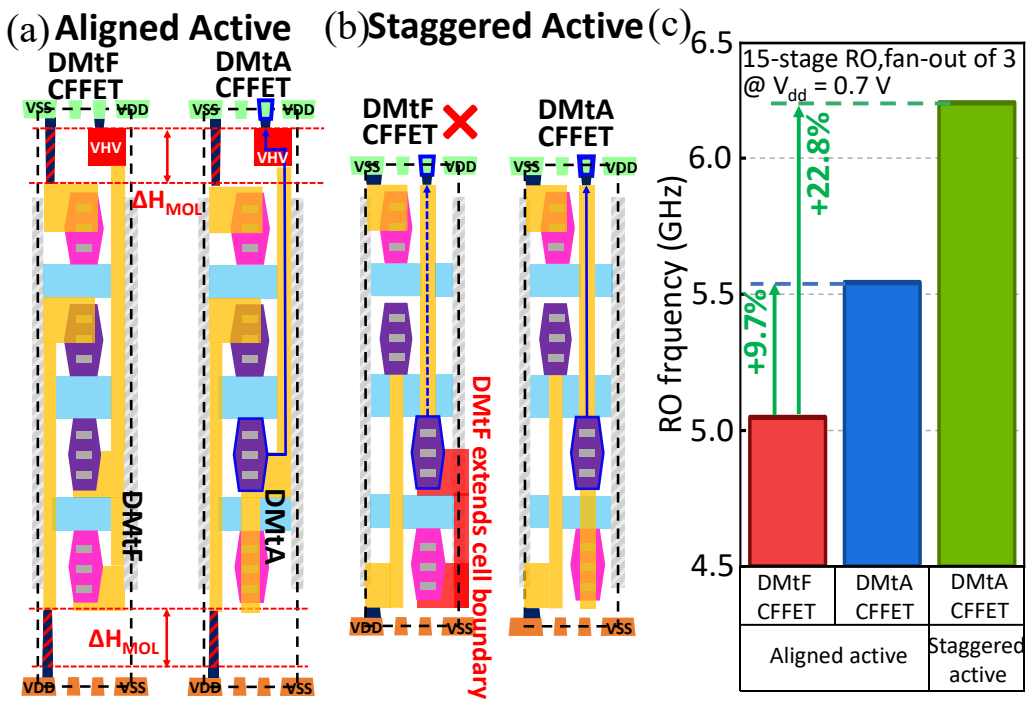

**Fig. 22** DMtA-enabled performance boosting in 4-tier-transistor CFFET [10]. (a) The VHV routing required by the aligned-active CFFET increases the MOL vertical-interconnect height ($\Delta H_{MOL}$) and resistance. (b) A staggered-active arrangement enables direct vertical connection without VHV, but only DMtA implements this scheme without extending the DM beyond the cell boundary and incurring an area penalty. (c) RO frequency comparison of aligned- and staggered-active DMtA CFFET against aligned-active DMtF CFFET.